# A Meshless-based Local Reanalysis Method for Structural Analysis

Zhenxing Cheng[a], Hu Wang[a*]

[a] *State Key Laboratory of Advanced Design and Manufacturing for Vehicle Body, Hunan University, Changsha, 410082, P.R. China*

**Abstract** This study presents a meshless-based local reanalysis (MLR) method. The purpose of this study is to extend reanalysis methods to the Kriging interpolation meshless method due to its high efficiency. In this study, two reanalysis methods: combined approximations CA) and indirect factorization updating (IFU) methods are utilized. Considering the computational cost of meshless methods, the reanalysis method improves the efficiency of the full meshless method significantly. Compared with finite element method (FEM) based reanalysis methods, the main superiority of meshless-based reanalysis method is to break the limitation of mesh connection. The meshless-based reanalysis is much easier to obtain the stiffness matrix $\mathbf{K}_m$ even for solving the mesh distortion problems. However, compared with the FEM-based reanalysis method, the critical challenge is to use much more nodes in the influence domain due to high order interpolation. Therefore, a local reanalysis method which only needs to calculate the local stiffness matrix in the influence domain is suggested to improve the efficiency further. Several typical numerical examples are tested and the performance of the suggested method is verified.

**Keywords** Reanalysis, Meshless, Kriging interpolation, Combined approximations, Indirect factorization updating

## 1 Introduction

Analysis, as an important issue of structural design and optimization, always must be repeated because design is a continuously modified process. For most of the complicated engineering problems, the computational cost is commonly high due to repeated full analysis. Therefore, efficient computation is urgently needed [1]. Many high performance computational methods have been developed, such as GPU/CPU computational methods [2, 3], parallel computation [4, 5], reanalysis methods [6] and so on. The reanalysis method has been presented to predict the response of modified structures efficiently without full analysis. In the past several decades, many reanalysis methods have been developed which can be divided into two

---

[*] Corresponding author Tel: +86 0731 88655012; fax: +86 0731 88822051.

E-mail address: wanghu@hnu.edu.cn (Hu Wang)

categories: direct methods (DMs) and approximate methods.

The direct methods are used to obtain the exact response of the structure with local or low-rank modifications. There are mainly two strategies of DMs: Sherman-Morrison-Woodbury (SMW) formula and matrix factorization updating method. The earliest Sherman-Morrison formula can only solve the problems of rank-one modification [7], then Woodbury developed it for multiple rank-one modification [8]. At this point, the SWM formula has become a feasible and effective DM for low-rank modification. Another popular DM is based on the matrix factorization updating strategy. It is also proposed to solve rank-one modification [9], and then extended for multiple-rank modifications [10]. Liu et al. applied this method for the structure reanalysis with added DOFs [11, 12]. Meanwhile, Song et al. extended the matrix factorization updating method to sparse matrix solution [13]. Moreover, many other direct methods have been developed recently. Huang and Wang presented an Independent Coefficient (IC) method to solve large-scale problems [14], and suggested an Indirect Factorization Updating (IFU) method for boundary modifications [15]. Gao and Wang presented an exact block-based reanalysis method for local modifications [16].

Compared with DMs, approximate methods can solve the high-rank modifications, but the exact response usually cannot be obtained. The CA method developed by Kirsch might be the most popular approximate reanalysis method due to its generality [6]. By combining with local and global approximation methods, the CA method inherited the efficiency of local approximation and the accuracy of global approximation. Initially, the classical CA method was only suitable for small modifications [17, 18]. Therefore, Kirsch and Papalambros presented a method to break the bottleneck by introducing a modified initial design (MID) [17]. Chen proposed an approximate two-step method to solve problems of adding DOFs [19]. Sequentially, the CA method has been applied to multiple disciplines, such as structural static reanalysis [6, 20], eigenvalue reanalysis [21-23], topological optimization [24, 25], vibration reanalysis [26], linear dynamic reanalysis [27, 28], nonlinear dynamic reanalysis [29, 30], sensitivity reanalysis [31, 32] and other fields[33]. Moreover, the reanalysis methods have been applied to some optimization applications. For example, Zuo et al. used reanalysis method to improve the efficiency of genetic algorithm (GA) for structural optimization [34]. Sun et al. proposed an adaptive reanalysis method for structural optimization [35]. To improve the efficiency of reanalysis method, a parallel CA method was first developed by Wang et al. [5]. Then

He et al. suggested a multiple-GPU based parallel IC reanalysis method which breaks through the memory bottleneck of GPU [4]. To extend the applications of reanalysis, Gao and Wang suggested an adaptive time-based global reanalysis (ATGR) method which is based on Newmark-$\beta$ method [36]. Materna et al. suggested a residual increment approximations for both linear and nonlinear reanalysis [37]. Kaveh et al. applied reanalysis to the near-regular mechanical systems [38-41]. Based on these techniques, Wang *et al.* developed a CAD/CAE integrated parallel reanalysis design system [42]. Recently, Huang and Wang proposed a multi-grid reanalysis method for re-meshed finite element models [43], and it is very suitable for the methods which need re-meshed the model, such as adaptive model order reduction methods [44-47].

Although the reanalysis method has been well developed, it is still difficult to apply in complicated models. In our opinions, the most critical bottleneck is how to obtain the change of stiffness matrix, because the meshes of initial and modified models should be identified. Compared with FEM, the meshless method is independent of meshes and the change of the stiffness matrix is easy to obtain, and considering the expensive computational cost of meshless method, the reanalysis method should accelerate it significantly. Therefore, we hope to extend the reanalysis for the meshless method.

Many meshless methods have been proposed in recent decades [48, 49]. The earliest meshless method is smoothed particle hydrodynamics (SPH) [50]. After that, many kinds of meshless methods have been proposed, such as diffuse element method (DEM) [51], element free Galerkin (EFG) method [52], reproducing kernel particle method (RKPM) [53], hp-meshless cloud method [54], meshless local Petrov–Galerkin method (MLPG) [55], point interpolation method [56], radial point interpolation method (RPIM) [57], moving Kriging meshless method [58, 59], efficient meshless method [60, 61]and several others. In this study, the moving Kriging (MK) meshless method is utilized. The MK meshless method is a kind of weak-form meshless methods which was developed on the EFG method, and the MK interpolation procedure is employed to replace the moving least squared (MLS) procedure [58]. This method behaves good stability and excellent accuracy due to its properties of weak-form [62, 63]. Kriging interpolation procedure is an optimal interpolation algorithm proposed by Matheron and Krige [64]. For meshless methods, Gu firstly introduced MK interpolation procedure in element free Galerkin (EFG) method [58], then Tongsuk and Kanok-Nukulchai applied this method to one and two-dimensional elasticity problems [65]. Sayakoummane and Kanok-Nukulchai extended this method to shell structures [66]. Moreover, Bui et al. largely expanded the applied range of the moving

Kriging method, such as plate structures [67-73], piezoelectric structures [74], and structural dynamic analysis [75, 76]. Shaw et al. applied the Kriging interpolation with an error-reproduction kernel method to solve linear and nonlinear boundary value problems [77]. The Kriging interpolation also has been extended to the fracture analysis [78, 79]. Instead of global formulations described previously, Lam et al. proposed a local weak-form meshless formulation incorporating Kriging-based shape functions to form a novel local Kriging meshless method for two-dimensional structural analysis [80] and it had been applied to dynamic nonlinear problems [81, 82]. Recently, the local Kriging method was employed to solve two-dimensional and three-dimensional transient heat conduction problems [83], elastodynamic analysis for two-dimensional solids [84], free vibration [85] and thermal bucking [86] analysis of functionally graded plates.

Compared with FEM, the pre-processing of meshless method is more convenient because the meshless method only needs the nodal information while information of meshes is needless. Furthermore, the meshless method behaves strong adaptability and it's easy to model, but usually the computational cost of meshless method is higher than FEM. Based on this point, the computational cost of meshless method will be significantly reduced by reanalysis method.

Therefore, a reanalysis method named meshless-based local reanalysis (MLR) method is suggested in this study. In this method, a structure should be analyzed by meshless method first. Then the MLR method should be used to predict the response of modified structure during design process. Compared with the FEM-based reanalysis method, it is easier to be implemented because only nodes should be added or removed while a structure is modified. Moreover, the suggested method is based on the MK meshless method which the Kriging interpolation is used to construct the shape function due to satisfying the property of Kronecker's delta function. Furthermore, a local search strategy for updating change of stiffness matrix is employed in this study to improve the efficiency.

The rest of this paper is represented as follows. Basic theories of the MK meshless method is introduced in Section 2. The MLR method and the local strategy are introduced in Section 3. In Section 4, four typical numerical examples are shown to test the performance of the MLR method. Finally, the conclusions are summarized in Section 5.

## 2 Basic theories

### 2.1 Framework of the MLR

The MLR method extended reanalysis methods to the MK meshless method due to its high efficiency, and a local reanalysis algorithm is suggested to improve the efficiency much more. The framework of the MLR method is presented in Fig. 1.

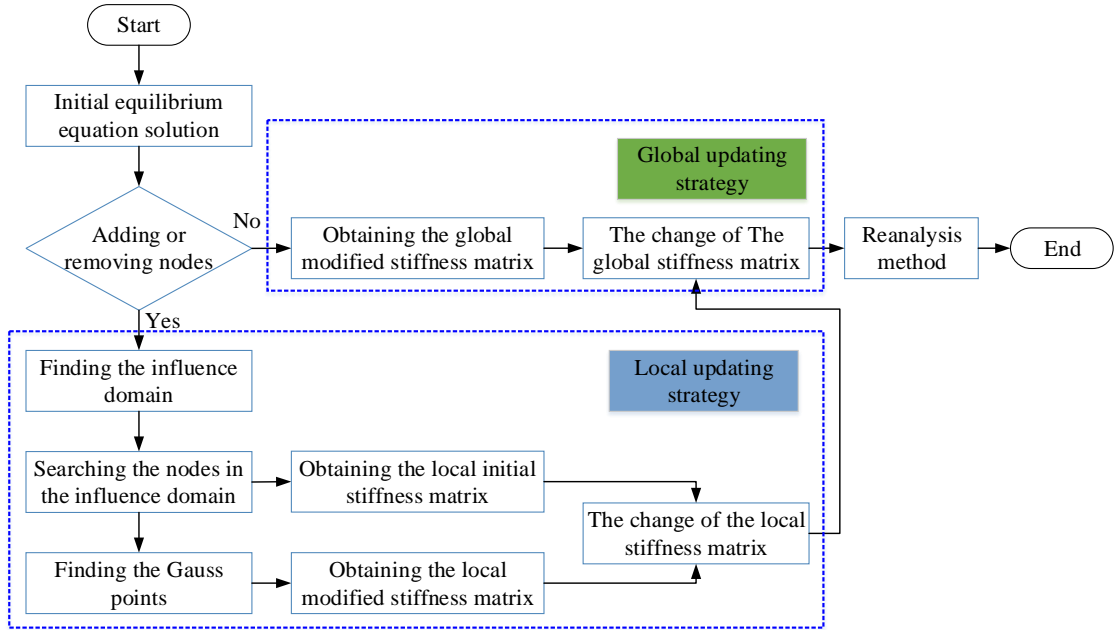

Fig. 1 Framework of the MLR method

It can be found that the framework of the MLR method is divided into two parts: global and local updating strategy. The global updating strategy needs to calculate the stiffness matrix of all nodes in solution domain while the local updating strategy only needs to calculate the stiffness matrix of the nodes in the influence domain. Generally, structural modifications can be divided into three parts: the number of DOFs is unchanged, decreased or increased [6]. It's important to note that, the local search strategy for updating stiffness matrix of MLR method is unavailable while solving the problems that number of DOFs is constant, such as change of material parameters (Young's modulus or Poisson's ratio). However, the global stiffness matrix can be obtained by the MK meshless method directly.

Thus, the MLR method is suitable for local modifications. However, for large or overall modifications, a global stiffness matrix updating strategy was suggested to solve such problems by calculating global stiffness matrix directly.

## 2.2 Moving Kriging meshless method

The MK meshless method has already been studied by several scholars, and the mathematical formulas can be found in their literature [58, 59]. Assuming the distribution functions $u(\mathbf{x}_i)$ in a sub-domain $\Omega_x$, so that $\Omega_x \subseteq \Omega$. Supposing that $u(\mathbf{x}_i)$ can be interpolated by nodal values $\mathbf{x}_i (i \in [1, n])$, $n$ is the

number of the nodes in $\Omega_x$. Define the MK interpolation $\mathbf{u}^h(\mathbf{x})$ as

$$\mathbf{u}^h(\mathbf{x}) = [\mathbf{p}^T(\mathbf{x})\mathbf{S}_a + \mathbf{r}^T(\mathbf{x})\mathbf{S}_b]\mathbf{u}(\mathbf{x}) \tag{1}$$

or

$$\mathbf{u}^h(\mathbf{x}) = \sum_I^n \varphi_I(\mathbf{x})u_I, \tag{2}$$

where $\varphi_I(\mathbf{x})$ means the MK shape function and it can be rewritten as

$$\varphi_I(\mathbf{x}) = \sum_j^m p_j(\mathbf{x})S_{a_{jI}} + \sum_k^n r_k(\mathbf{x})S_{b_{kI}}, \tag{3}$$

where $S_{a_{jI}}$ denotes the element of matrix $\mathbf{S}_a$ at row $j$ and column $I$, $S_{b_{kI}}$ denotes the element of matrix $\mathbf{S}_b$ at row k and column $I$. Matrix $\mathbf{S}_a$ and matrix $\mathbf{S}_b$ are defined by the following equations:

$$\mathbf{S}_a = (\mathbf{P}^T\mathbf{R}^{-1}\mathbf{P})^{-1}\mathbf{P}^T\mathbf{R}^{-1}, \tag{4}$$

$$\mathbf{S}_b = \mathbf{R}^{-1}(\mathbf{I} - \mathbf{P}\mathbf{S}_a), \tag{5}$$

where $\mathbf{I}$ is a unit matrix and the matrix $\mathbf{P}$ is defined as:

$$\mathbf{P} = \begin{bmatrix} p_1(\mathbf{x}_1) & p_2(\mathbf{x}_1) & \cdots & p_m(\mathbf{x}_1) \\ p_1(\mathbf{x}_2) & p_2(\mathbf{x}_2) & \cdots & p_m(\mathbf{x}_2) \\ \vdots & \vdots & \ddots & \vdots \\ p_1(\mathbf{x}_m) & p_2(\mathbf{x}_m) & \cdots & p_m(\mathbf{x}_m) \end{bmatrix}. \tag{6}$$

Moreover, $\mathbf{r}(\mathbf{x})$ in Eq.(1) can be defined by

$$\mathbf{r}(\mathbf{x}) = \{R(\mathbf{x}_1, \mathbf{x}) \quad R(\mathbf{x}_2, \mathbf{x}) \quad \cdots \quad R(\mathbf{x}_n, \mathbf{x})\}^T. \tag{7}$$

$R(\mathbf{x}_i, \mathbf{x}_j)$ is the correlation function between any pair of nodes $\mathbf{x}_i$ and $\mathbf{x}_j$, it can be calculated by

$$R(\mathbf{x}_i, \mathbf{x}_j) = \text{cov}[u(\mathbf{x}_i)u(\mathbf{x}_j)]. \tag{8}$$

Many functions could be chosen as a correlation function [58]. A widely used Gaussian function is employed, and it can be calculated by

$$R(\mathbf{x}_i, \mathbf{x}_j) = e^{-\theta r_{ij}^2}, \tag{9}$$

where $r_{ij} = \|\mathbf{x}_i - \mathbf{x}_j\|$, in which $\theta$ is a correlation parameter. It has been studied thoroughly by Bui [59, 72]. In this study, we didn't do much more research in the correlation parameter, but a suitable value of the correlation parameter is given as $\theta = 1$ for all the cases by referring to [59]. The linear basic $\mathbf{p}^T(\mathbf{x}) = [1 \quad x \quad y]$ is using for numerical analysis. In addition, matrix $\mathbf{R}[R(\mathbf{x}_i, \mathbf{x}_j)]_{n \times n}$ can be defined by

$$\mathbf{R}[R(\mathbf{x}_i, \mathbf{x}_j)] = \begin{bmatrix} 1 & R(\mathbf{x}_1, \mathbf{x}_2) & \cdots & R(\mathbf{x}_1, \mathbf{x}_n) \\ R(\mathbf{x}_2, \mathbf{x}_1) & 1 & \cdots & R(\mathbf{x}_2, \mathbf{x}_n) \\ \vdots & \vdots & \ddots & \vdots \\ R(\mathbf{x}_n, \mathbf{x}_1) & R(\mathbf{x}_n, \mathbf{x}_2) & \cdots & 1 \end{bmatrix}. \tag{10}$$

In many problems, the first-order derivative is required, and it can be obtained from

$$\varphi_{I,i}(\mathbf{x}) = \sum_j^m p_{j,i}(\mathbf{x})S_{a_{jI}} + \sum_k^n r_{k,i}(\mathbf{x})S_{b_{kI}}. \tag{11}$$

For almost all static problems, usually can be simplified as an equilibrium equation:

$$\mathbf{K}_m \mathbf{U} = \mathbf{F} \tag{12}$$

where $\mathbf{K}_m$, $\mathbf{F}$, $\mathbf{U}$ mean the stiffness matrix, the load vector, the unknown displacement vector

respectively, and the stiffness matrix $\mathbf{K}_m$ can be defined as

$$\mathbf{K}_m = \begin{bmatrix} \mathbf{K}_{11} & \mathbf{K}_{12} & \cdots & \mathbf{K}_{1n} \\ \mathbf{K}_{21} & \mathbf{K}_{22} & \cdots & \mathbf{K}_{2n} \\ \vdots & \vdots & \ddots & \vdots \\ \mathbf{K}_{n1} & \mathbf{K}_{n2} & \cdots & \mathbf{K}_{nn} \end{bmatrix} \quad (13)$$

where

$$\mathbf{K}_{ij} = \int_\Omega \mathbf{B}_i^T \mathbf{D} \mathbf{B}_j d\Omega \quad (i,j=1,2,\cdots,n). \quad (14)$$

In the Eq.(14), $\mathbf{D}$ is the constitutive matrix of material and $\mathbf{B}_i$, $\mathbf{B}_j$ can be calculated by

$$\mathbf{B}_i = \begin{bmatrix} \varphi_{i,x} & 0 \\ 0 & \varphi_{i,y} \\ \varphi_{i,y} & \varphi_{i,x} \end{bmatrix} \text{ (2D formula)} \quad (15)$$

or

$$\mathbf{B}_i = \begin{bmatrix} \varphi_{i,x} & 0 & 0 \\ 0 & \varphi_{i,y} & 0 \\ 0 & 0 & \varphi_{i,z} \\ 0 & \varphi_{i,z} & \varphi_{i,y} \\ \varphi_{i,z} & 0 & \varphi_{i,x} \\ \varphi_{i,y} & \varphi_{i,x} & 0 \end{bmatrix} \text{ (3D formula)}, \quad (16)$$

where $\varphi_{i,x}$, $\varphi_{i,y}$, $\varphi_{i,z}$ can obtained from Eq.(11). Then the stiffness sub-matrix can be calculated by Eq.(14), then assemble the sub-matrix into a global stiffness matrix.

In order to obtain the integral in Eq.(14), a background cell structure which is independent of the nodes should be used. In each background cell, the $2\times 2$ Gauss quadrature is used as shown in Fig. 2. An important issue in the MK meshless method is the definition of support domain. It is defined to determine how many discrete nodes is in the interpolated domain. There is no method which can totally suitably determine all types of nodal distributions actually, but the accuracy of the method deeply relies on the number of nodes inside the influence domain. The influence domain usually can be defined as a circular region with a Gauss integration point as its center, and it is shown in Fig. 2. Normally, the size of the support domain is evaluated by the following formula:

$$d_m = \alpha d_c \quad (17)$$

where $d_c$ is defined as the distance between adjacent nodes which is locally around the point of interest, and the factor $\alpha$ is a scale coefficient. In this study $\alpha = 3$ has been used by referring to the literatures [52, 87].

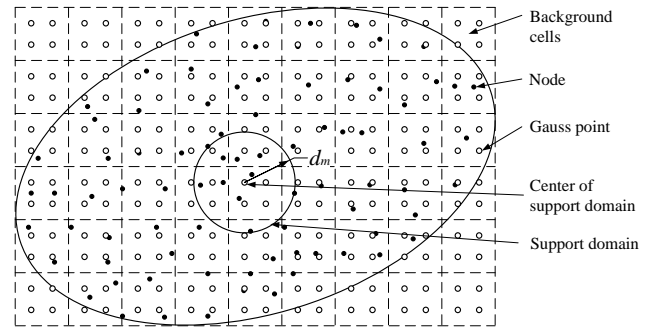

Fig. 2 Definition of support domain in the MK meshless method

## 3 Meshless-based Local Reanalysis method

In this study, two reanalysis algorithms, CA and IFU, have been used to improve the computational efficiency. In order to improve the efficiency much more, a local reanalysis algorithm is suggested. The framework of MLR is shown in Fig. 1. In this section, the detail of the MLR is expounded.

## 3.1 Local search strategy for updating changed stiffness matrix

The MLR method integrates the reanalysis and MK meshless methods seamlessly. The MK meshless method is used to calculate the initial response $\mathbf{U}^*$ by solving Eq.(21) and obtain the modified stiffness matrix $\mathbf{K}_m$. Then the modified response $\mathbf{U}$ can be predicted by reanalysis methods.

A local search strategy for updating changed stiffness matrix is suggested to improve the efficiency, and only the nodes inside the influence domain are used to construct the local stiffness matrix by this strategy. The influence domain can be defined by background cells, usually the influence domain includes the background cells and their contiguous background cells where the modification located as shown in Fig. 3. Obviously, there are four background cells associated with the modification. Therefore, the influence domain is composed of four background cells and their contiguous background cells. Similarly, the strategy is also suitable for problems of adding nodes, as shown in Fig. 4. When some nodes are added to the solution domain, there are four background cells associated with the modification too. Then, the influence domain is also composed of four background cells and their contiguous background cells.

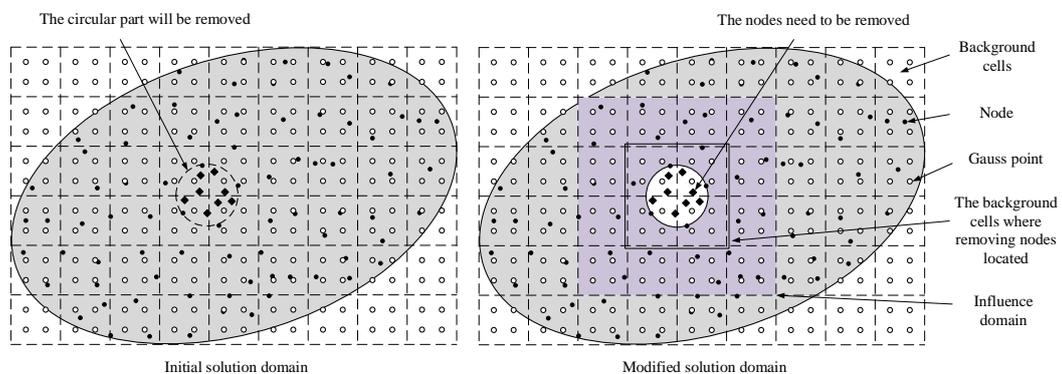

Fig. 3 The influence domain caused by removing nodes

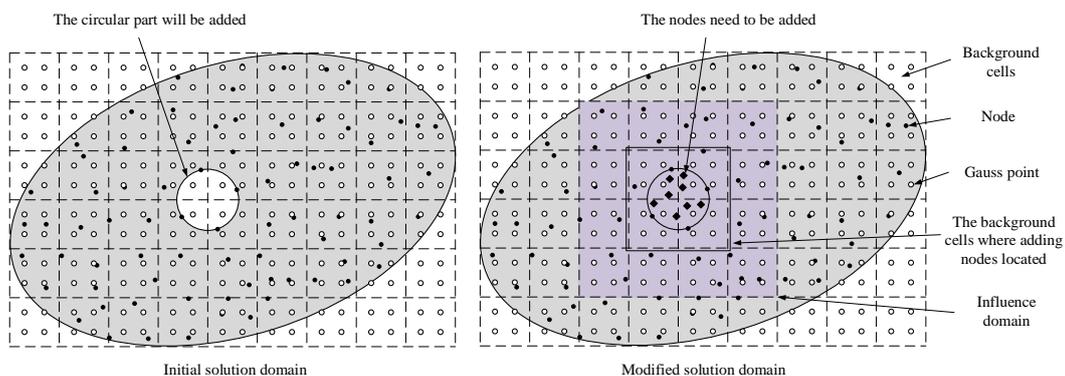

Fig. 4 The influence domain caused by adding nodes

While the initial stiffness matrix $\mathbf{K}_m^*$ and the initial displacement vector $\mathbf{U}^*$ have been obtained by the MK meshless method, the key issue of reanalysis method is how to obtain the modified stiffness matrix $\mathbf{K}_m$. Usually, the modified stiffness matrix $\mathbf{K}_m$ can be calculated by the MK meshless method directly, then $\Delta\mathbf{K}_m$ can be obtained by Eq.(23). In the FEM-based reanalysis, only the nodes related the modified elements need to be considered. However, due to high order interpolation and influence domain, all nodes in the influence domain of changed nodes should be found and involved in building. Fortunately, for most of the structural design and optimization, the change is local.

Therefore, a local search strategy has been suggested to obtain the change of stiffness matrix $\Delta\mathbf{K}_m$, and it only needs to calculate the stiffness of nodes inside the influence domain rather than all nodes. The local search strategy for updating changed stiffness matrix can save much computational cost and improve the efficiency in the problems of small changes because it only needs to calculate a small part of the entire structure. As shown in Fig. 1, the steps using to obtain $\Delta\mathbf{K}_m$ and $\mathbf{K}_m$ by the MLR method can be summarized as following:

(1) Searching the nodes which have been added or removed and recording the serial number of nodes. For example, as shown in Fig. 3, the circular part should be removed after modification, so the nodes inside circle should be searched in this step.

(2) Finding the background cells which adding or removing nodes are located on, and constructing the influence domain.

(3) Searching the nodes in the influence domain. Then the local initial stiffness matrix $\mathbf{K}_{m_L}^*$ can be isolated from the global initial stiffness matrix $\mathbf{K}_m^*$.

(4) Finding the Gauss points which are associated with all the recorded nodes, then the local modified stiffness matrix $\mathbf{K}_{m_L}$ can be obtained by the MK method, and the change of local stiffness matrix $\Delta\mathbf{K}_{m_L}$ can be obtained from

$$\Delta\mathbf{K}_{m_L} = \mathbf{K}_{m_L} - \mathbf{K}_{m_L}^*. \qquad (18)$$

(5) Obviously, because the change of stiffness matrix of outside influence domain is zero, the change of global stiffness matrix is equal to the change of local stiffness matrix.

$$\Delta\mathbf{K}_m = \Delta\mathbf{K}_{m_L} \qquad (19)$$

(6) The global modified stiffness matrix can be obtained from

$$\mathbf{K}_m = \mathbf{K}_m^* + \Delta\mathbf{K}_m. \qquad (20)$$

**3.2 Meshless-based CA reanalysis**

Assuming the equilibrium equation of initial structure is given by the following equation:

$$\mathbf{K}_m^* \mathbf{U}^* = \mathbf{F},\tag{21}$$

where $\mathbf{K}_m^*$ means the stiffness matrix which can be obtained by the MK meshless method from Eq.(13), $\mathbf{U}^*$, $\mathbf{F}$ denote the displacement vector and the load vector respectively. After redesigning the structure, the modified equilibrium equation changed as

$$\mathbf{K}_m \mathbf{U} = \mathbf{F},\tag{22}$$

where

$$\mathbf{K}_m = \mathbf{K}_m^* + \Delta \mathbf{K}_m.\tag{23}$$

Obviously, the displacement vector $\mathbf{U}$ will be predicted by reanalysis method rather than full analysis. Assuming that the displacements $\mathbf{U}$ of a new design can be estimated by a linear combination of $s$ independent basis vectors, $\mathbf{U}_1, \mathbf{U}_2, \cdots, \mathbf{U}_s$:

$$\mathbf{U} = y_1 \mathbf{U}_1 + y_2 \mathbf{U}_2 + \cdots + y_s \mathbf{U}_s = \mathbf{U}_B \mathbf{y}.\tag{24}$$

Assuming that $s \ll n$, $\mathbf{U}_B$ is the $n \times s$ matrix of the basis vectors and $\mathbf{y}$ is a vector of coefficients to be determined:

$$\mathbf{U}_B = [\mathbf{U}_1, \mathbf{U}_2, \cdots, \mathbf{U}_s],\tag{25}$$

$$\mathbf{y}^T = [y_1, y_2, \cdots, y_s].\tag{26}$$

The response $\mathbf{U}$ can be predicted by the following steps:

(1) Constructing the matrix of the basis vectors $\mathbf{U}_B$ by Eq.(25). The initial value and the following series of basis vectors can be obtained from:

$$\mathbf{U}_1 = (\mathbf{K}_m^*)^{-1} \mathbf{F} = \mathbf{U}^*,\tag{27}$$

$$\mathbf{U}_{i+1} = -\mathbf{B}\mathbf{U}_i \quad (i = 1, 2, \cdots, s),\tag{28}$$

where the matrix $\mathbf{B}$ is defined as

$$\mathbf{B} = (\mathbf{K}_m^*)^{-1} \Delta \mathbf{K}_m.\tag{29}$$

(2) Constructing the reduced stiffness matrix $\mathbf{K}_R$ and load vector $\mathbf{F}_R$ by:

$$\mathbf{K}_R = \mathbf{U}_B^T \mathbf{K}_m \mathbf{U}_B, \quad \mathbf{F}_R = \mathbf{U}_B^T \mathbf{F}.\tag{30}$$

(3) Calculating the vector of coefficients $\mathbf{y}$ by solving the following equation:

$$\mathbf{K}_R \mathbf{y} = \mathbf{F}_R.\tag{31}$$

(4) Updating the modified displacement by Eq.(24).

(5) In order to calculate the accuracy of stress-strain results, the modified strain and stress should be calculated by the following equations:

$$\boldsymbol{\varepsilon} = \begin{bmatrix} \dfrac{\partial}{\partial x} & 0 \\ 0 & \dfrac{\partial}{\partial y} \\ \dfrac{\partial}{\partial y} & \dfrac{\partial}{\partial x} \end{bmatrix} \mathbf{U}_B \mathbf{K}_R^{-1} \mathbf{F}_R,\tag{32}$$

$$\boldsymbol{\sigma} = \mathbf{c}\boldsymbol{\varepsilon},\tag{33}$$

where $\mathbf{c}$ is the matrix of material constants obtained by experiments.

### 3.3 Meshless-based IFU reanalysis

Compared with the CA method, the IFU method only calculates the displacement of the influenced DOFs when solving Eq.(22). Assuming the modified displacement is

$$\mathbf{U} = \mathbf{U}^* + \Delta \mathbf{U},\tag{34}$$

then rewrite Eq.(22) as

$$\mathbf{K}_m(\mathbf{U}^* + \Delta\mathbf{U}) = \mathbf{F} \tag{35}$$

or

$$\mathbf{K}_m \Delta\mathbf{U} = \mathbf{F} - \mathbf{K}_m \mathbf{U}^*. \tag{36}$$

Defining the residual value of the initial displacement $\boldsymbol{\delta}$ as

$$\boldsymbol{\delta} = \mathbf{F} - \mathbf{K}_m \mathbf{U}^*. \tag{37}$$

It worth mentioning that only some members of $\boldsymbol{\delta}$ are non-zero while the modification is local. Basing on this property, the modified displacement $\mathbf{U}$ can be predicted by the following steps:

(1) Calculating the Cholesky factorization of initial stiffness matrix by Eq.(38);

$$\mathbf{K}_m^* = \mathbf{L}_0 \mathbf{L}_0^T \tag{38}$$

(2) Calculating measurement vector: $\boldsymbol{\delta}$ and $\boldsymbol{\Delta}$ by Eq.(37) and Eq.(39) respectively;

$$\boldsymbol{\Delta} = sum\left(\left|\mathbf{K}_m - \mathbf{K}_m^*\right|\right) + |\boldsymbol{\delta}| \tag{39}$$

(3) Recording the unbalanced DOFs: If $|\Delta(i)| > 0$, the $i$-th DOF is unbalanced, and $i$ should be recorded in $\mathbf{S}_d$ by ascending order, and the number of unbalanced DOFs is $n_d$.

(4) Extracting unbalanced equations:

For $i=1$ to $n_d$

$\mathbf{K}_u(i,:) = \mathbf{K}_m(\mathbf{S}_d(i),:)$ ; $\boldsymbol{\delta}_u(i) = \boldsymbol{\delta}(\mathbf{S}_d(i))$ ;

End for.

(5) Applying extra constrains on $\mathbf{K}_m^*$:

For $i = n_d$ to 1

$\mathbf{V}(:,i) = \mathbf{L}_0(:,\mathbf{S}_d(i))$ ; $\mathbf{V}(\mathbf{S}_d(i),i) = 0$ ;

$\mathbf{L}_0(\mathbf{S}_d(i),:) = \mathbf{0}; \mathbf{L}_0(:,\mathbf{S}_d(i)) = \mathbf{0}; \mathbf{L}_0(\mathbf{S}_d(i),\mathbf{S}_d(i)) = 1$ ;

End for.

(6) Calculating the right-hand vectors or extra constraints $\mathbf{R}$:

For $i=1$ to $n_d$

$\mathbf{R}(:,i) = \mathbf{K}_m(:,\mathbf{S}_d(i))$; $\mathbf{R}(\mathbf{S}_d(i),:) = \mathbf{0}$; $\mathbf{R}(\mathbf{S}_d(i),i) = 1$;

End for.

(7) Calculating the fundamental solution system $\mathbf{B}$ of the balanced equations by solving the Eq.(40) using SWM formula [8].

$$(\mathbf{L}_0 \mathbf{L}_0^T + \mathbf{V}\mathbf{V}^T)\mathbf{B} = \mathbf{R} \tag{40}$$

(8) Reducing the unbalanced equations by $\mathbf{K}_R = \mathbf{K}_u \mathbf{B}$.

(9) Solving the reduced equation by $\mathbf{K}_R \mathbf{y} = \boldsymbol{\delta}_u$.

(10) Calculating the increment of displacements by $\Delta\mathbf{U} = \mathbf{B}\mathbf{y}$.

(11) Calculating modified displacement $\mathbf{U}$ by Eq.(34).

(12) The modified strain and stress can be calculated by Eq.(41) and Eq.(42) to test the accuracy of stress-strain results.

$$\boldsymbol{\varepsilon} = \begin{bmatrix} \dfrac{\partial}{\partial x} & 0 \\ 0 & \dfrac{\partial}{\partial y} \\ \dfrac{\partial}{\partial y} & \dfrac{\partial}{\partial x} \end{bmatrix} (\mathbf{U}^* + (\mathbf{L}_0\mathbf{L}_0^T + \mathbf{V}\mathbf{V}^T)^{-1}\mathbf{R}\mathbf{K}_R\boldsymbol{\delta}_u),$$

(41)

$$\boldsymbol{\sigma} = \mathbf{c}\boldsymbol{\varepsilon}, \quad (42)$$

where **c** is the matrix of material constants obtained by experiments.

## 4 Numerical examples

In order to test the accuracy and efficiency of the MLR method, four examples will be solved by the proposed methods. These four cases involve 2D and 3D, decreased and increased DOFs, concentrated and uniformed load problems, thus the performance of the MLR method could be verified thoroughly. In this study, the comparison has been made between the MLR and full analysis, and the errors of displacement, Von Mises stress and Von Mises strain are defined by the following formulas:

$$E_u = \frac{\|\mathbf{U}_{MLR} - \mathbf{U}_{FA}\|}{\|\mathbf{U}_{FA}\|} \times 100\% \quad (43)$$

$$E_\varepsilon = \frac{\|\boldsymbol{\varepsilon}_{MLR} - \boldsymbol{\varepsilon}_{FA}\|}{\|\boldsymbol{\varepsilon}_{FA}\|} \times 100\% \quad (44)$$

$$E_\sigma = \frac{\|\boldsymbol{\sigma}_{MLR} - \boldsymbol{\sigma}_{FA}\|}{\|\boldsymbol{\sigma}_{FA}\|} \times 100\% \quad (45)$$

where $\mathbf{U}_{MLR}$, $\boldsymbol{\varepsilon}_{MLR}$, $\boldsymbol{\sigma}_{MLR}$ mean the results of MLR method, and $\mathbf{U}_{FA}$, $\boldsymbol{\varepsilon}_{FA}$, $\boldsymbol{\sigma}_{FA}$ mean the results of full analysis.

### 4.1 The rectangular plate optimization

As shown in the left of Fig. 5, a rectangular plate is considered. The dimension of the plate is $L \times D$ and the state of plane stress is considered, where the $L$=100$mm$, $D$=50$mm$, the Young's modulus $E$=200$GPa$, the Poisson's ratio $v$=0.3, and the vertical concentrated load $F$=1000$mN$. In order to obtain a lightweight design, the design of rectangular plate should be modified repeatedly by topology optimization method or others. Assuming the right of Fig. 5 is one of the modified designs. Then the response of modified structure will be predicted by the MLR method.

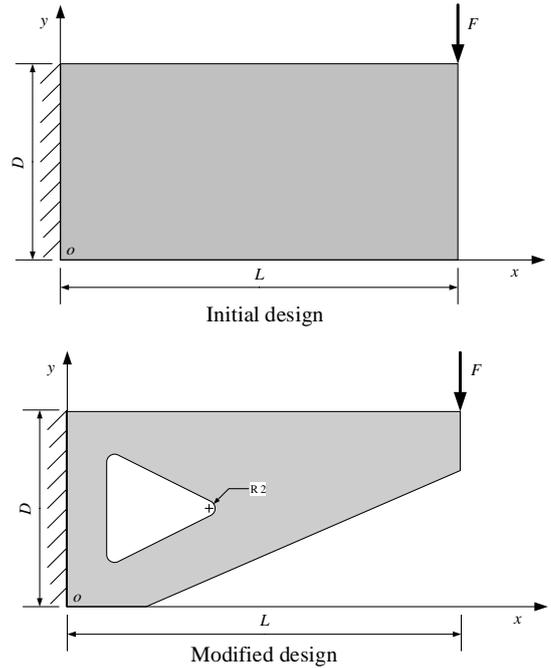

Fig. 5 The initial design and the modified structure of the rectangular plate

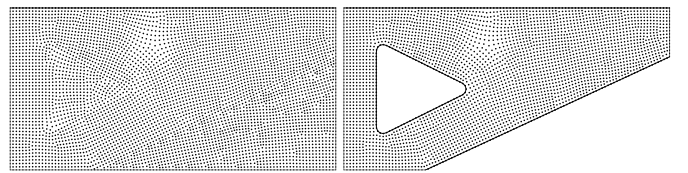

Fig. 6 The initial and the modified nodal distribution of the rectangular plate

Fig. 7 The error of displacement at different number of basis vectors

As shown in the left of Fig. 6, 5083 irregularly distributed field nodes are used in the initial structure. To investigate the performance of the MLR, the structure shown in the right of Fig. 6 is assumed to be the modified shape. The modified structure is only composed of 3341 nodes, and the percent of reduced DOFs is up to 34.3% of the original meshless model.

In general, the accuracy of the CA method relies on the number of basis vectors. Fig. 7 shows the variation in error of displacement when the number of basis vectors is increased. Obviously, the accuracy of the result is improved when the number of basis vectors is increased.

According to Fig. 7, it can be found that the accuracy of reanalysis solution converges when the number of basis vectors grows up to 10. Sequentially, with more basis vectors, the accuracy of predicted response can't be significantly improved. Moreover, analysis results comparisons between the MLR and the full analysis are illustrated in Fig. 8, Fig. 9 and Fig. 10, respectively. It is obvious that the result of the MLR is very close to the result of full analysis, and the displacements of some selected DOFs are listed in Tab.1. The displacement, strain and stress errors of the CA method with 10 basis vectors calculated by Eq.(43) are 0.075%, 0.69%, 0.61% respectively. However, all the errors of the IFU method are 0, it obtained the exact solutions.

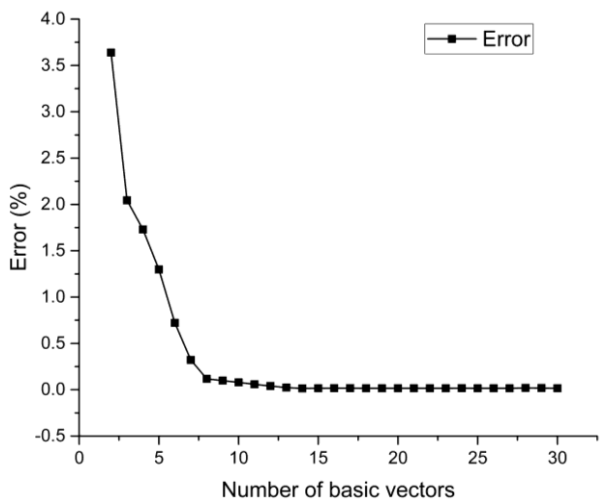

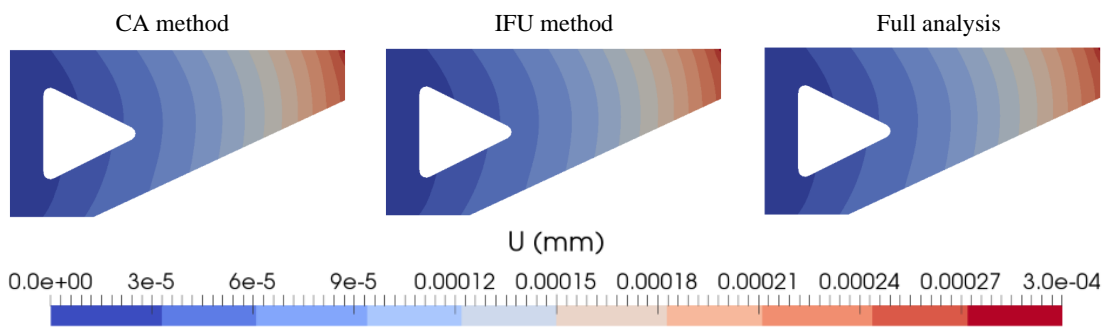

Fig. 8 The displacement results of the MLR and the full analysis methods

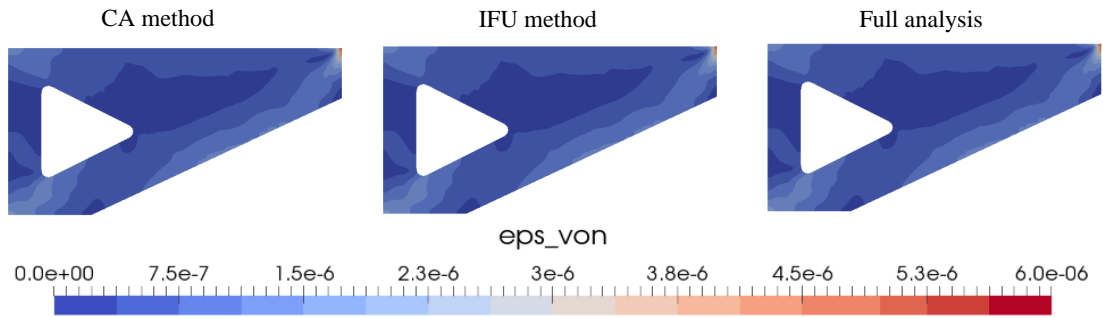

Fig. 9 The von Mises strain results of the MLR and the full analysis methods

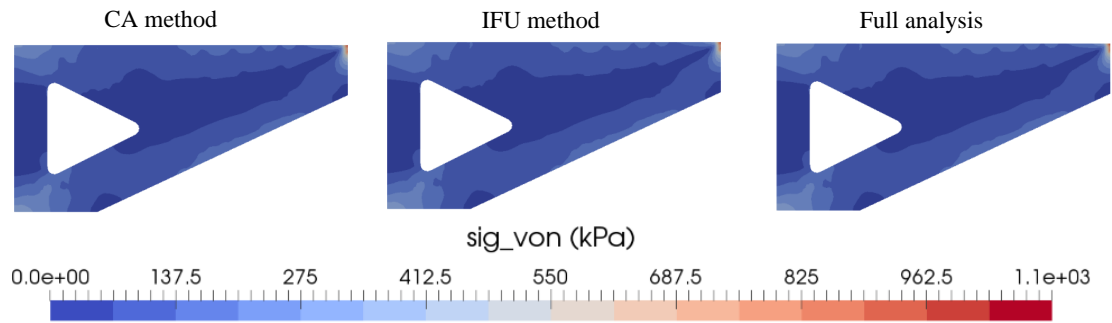

Fig. 10 The von Mises stress results of the MLR and the full analysis methods

Tab. 1 Displacement error analysis of the MLR method

| DOF ID | CA method | IFU method | Full analysis | Displacement error | |
|---:|---:|---:|---:|---:|---:|
| | | | | CA | IFU |
| 521 | -1.48859E-5 | -1.48728E-5 | -1.48728E-5 | 8.77844E-4 | 0 |
| 522 | -1.08025E-5 | -1.07985E-5 | -1.07985E-5 | 3.67652E-4 | 0 |
| 1111 | -9.52E-8 | -9.66E-8 | -9.66E-8 | 0.01395 | 0 |
| 1112 | -2.4733E-6 | -2.4784E-6 | -2.4784E-6 | 0.00203 | 0 |
| 2221 | 2.2155E-6 | 2.2152E-6 | 2.2152E-6 | 1.36715E-4 | 0 |
| 2222 | -1.0379E-6 | -1.037E-6 | -1.037E-6 | 8.98309E-4 | 0 |
| 5101 | 2.68836E-5 | 2.69254E-5 | 2.69254E-5 | 0.00155 | 0 |
| 5102 | -9.43354E-5 | -9.42683E-5 | -9.42683E-5 | 7.11762E-4 | 0 |

Tab. 2 Von Mises strain error analysis of the MLR method

| NODE ID | CA method | IFU method | Full analysis | Von Mises strain error | |
|---:|---:|---:|---:|---:|---:|
| | | | | CA | IFU |
| 100 | 1.8655E-6 | 1.8726E-6 | 1.8726E-6 | 0.00378 | 0 |
| 115 | 7.072E-7 | 6.995E-7 | 6.995E-7 | 0.01111 | 0 |
| 197 | 1.141E-6 | 1.1415E-6 | 1.1415E-6 | 4.45611E-4 | 0 |
| 369 | 1.1053E-6 | 1.1161E-6 | 1.1161E-6 | 0.00968 | 0 |
| 442 | 5.564E-7 | 5.587E-7 | 5.587E-7 | 0.00421 | 0 |
| 884 | 5.942E-7 | 5.919E-7 | 5.919E-7 | 0.00387 | 0 |
| 1057 | 6.34E-7 | 6.369E-7 | 6.369E-7 | 0.00457 | 0 |
| 2141 | 1.743E-7 | 1.754E-7 | 1.754E-7 | 0.00653 | 0 |

Tab. 3 Von Mises stress error analysis of the MLR method

| NODE ID | CA method | IFU method | Full analysis | Von Mises stress error | |
|---|---|---|---|---|---|
| | | | | CA | IFU |
| 100 | 370.25767 | 372.00852 | 372.00852 | 0.00471 | 0 |
| 115 | 155.78426 | 154.03874 | 154.03874 | 0.01133 | 0 |
| 197 | 251.40034 | 251.54859 | 251.54859 | 5.89379E-4 | 0 |
| 369 | 225.92672 | 227.88026 | 227.88026 | 0.00857 | 0 |
| 442 | 108.27899 | 106.57767 | 106.57767 | 0.01596 | 0 |
| 884 | 98.82425 | 99.51534 | 99.51534 | 0.00694 | 0 |
| 1057 | 135.48721 | 137.05081 | 137.05081 | 0.01141 | 0 |
| 2141 | 19.50026 | 20.57304 | 20.57304 | 0.05215 | 0 |

### 4.2 Support bracket redesign

A redesign of support bracket was considered [88]. The idea is to design a support bracket which will act as a cantilever beam to support an end load and will be fixed on two pin holes. Fig. 11 shows the geometry of the initial design where the fillet radius is 2.5*mm*. In order to reduce the stress concentration in the filler, the fillet radius is changed to 7.5*mm* in the modified design as shown in Fig. 11. A concentrated load *F* of 1000*mN* is applied at the free end, and the Young's modulus *E*=200*GPa*, the Poisson's ratio $\nu$=0.3.

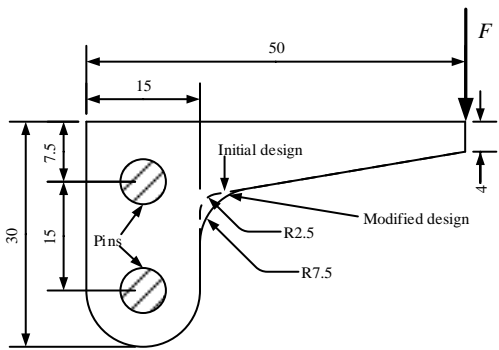

Fig. 11 The initial design and the modified design of the support bracket

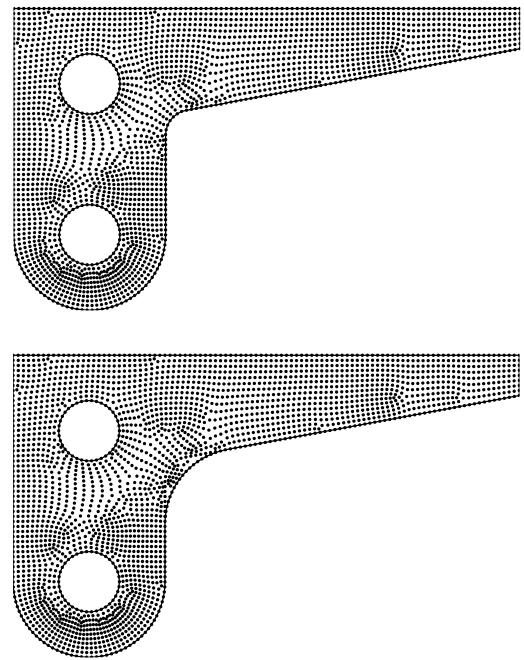

Fig. 12 The initial and the modified nodal distribution of the support bracket

As shown in the Fig. 12, 2611 irregularly distributed field nodes are used in the initial structure while only 2647 nodes are used in the modified structure, and the percent of adding DOFs is 1.4% of the original meshless model. Then the structural response will be predicted by the MLR method.

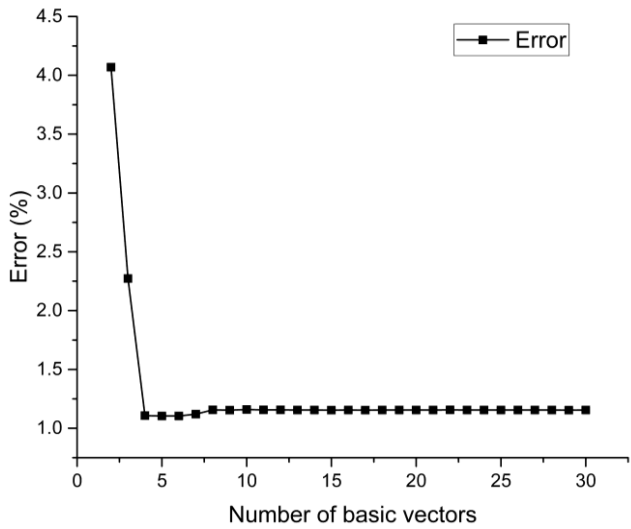

Fig. 13 The error of displacement at different number of basis vectors

basis vectors grows up to 8. Sequentially, with more basis vectors, the accuracy of predicted response can't be significantly improved. Moreover, analysis results comparisons between the MLR and the full analysis are illustrated in Fig. 14, Fig. 15 and Fig. 16, receptively. It is obvious that the result of the MLR is very close to the result of the full analysis, and the displacements of some selected DOFs are listed in Tab. 4. The displacement, strain and stress errors of the CA method are 1.1%, 6.7%, 5.1% respectively while all the errors of the IFU method are 0.

As shown in Fig. 13, it can be found that the accuracy of reanalysis solution converges when the number of

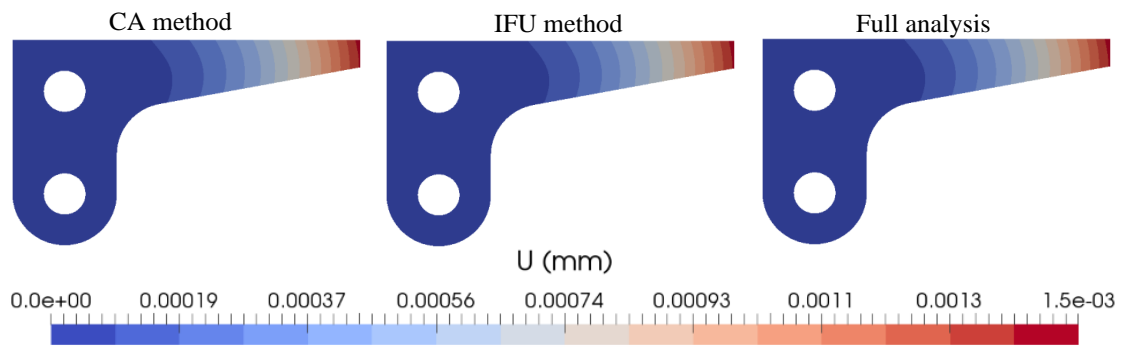

Fig. 14 The displacement result of the MLR method and the full analysis methods

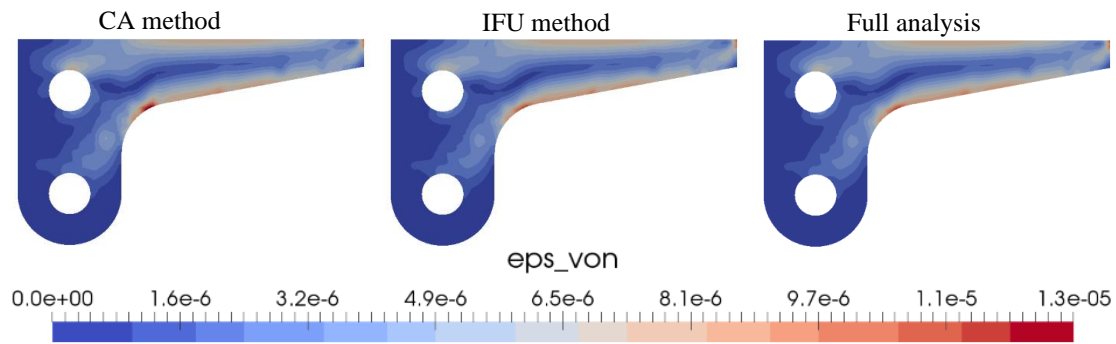

Fig. 15 The von Mises strain result of the MLR and the full analysis methods

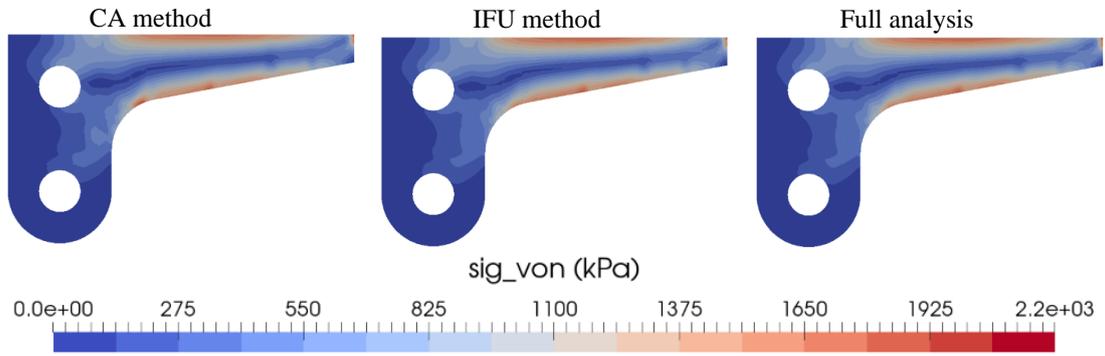

Fig. 16 The von Mises stress result of the MLR and the full analysis methods

Tab. 4 Displacement error analysis of the MLR method

| DOF ID | CA method | IFU method | Full analysis | Displacement error | |
|---|---|---|---|---|---|
| | | | | CA | IFU |
| 282 | -7.6368E-4 | -7.6032E-4 | -7.6032E-4 | 0.00442 | 0 |
| 283 | -6.86184E-5 | -7.09861E-5 | -7.09861E-5 | 0.03335 | 0 |
| 897 | 2.61588E-4 | 2.64118E-4 | 2.64118E-4 | 0.00958 | 0 |
| 898 | -0.00126 | -0.00127 | -0.00127 | 0.00347 | 0 |
| 2797 | 6.1366E-5 | 6.07579E-5 | 6.07579E-5 | 0.01001 | 0 |
| 2798 | -6.91991E-4 | -6.87641E-4 | -6.87641E-4 | 0.00633 | 0 |
| 3199 | 3.99416E-5 | 3.89738E-5 | 3.89738E-5 | 0.02483 | 0 |
| 3200 | -9.057E-4 | -9.04355E-4 | -9.04355E-4 | 0.00149 | 0 |

Tab. 5 Von Mises strain error analysis of the MLR method

| NODE ID | CA method | IFU method | Full analysis | Von Mises strain error | |
|---|---|---|---|---|---|
| | | | | CA | IFU |
| 451 | 2.9325E-6 | 3.0135E-6 | 3.0135E-6 | 0.02689 | 0 |
| 452 | 3.253E-6 | 3.3428E-6 | 3.3428E-6 | 0.02689 | 0 |
| 494 | 7.005E-6 | 7.2029E-6 | 7.2029E-6 | 0.02747 | 0 |
| 495 | 6.9909E-6 | 7.1869E-6 | 7.1869E-6 | 0.02726 | 0 |
| 817 | 4.93E-6 | 5.0274E-6 | 5.0274E-6 | 0.01938 | 0 |
| 818 | 4.9525E-6 | 5.0377E-6 | 5.0377E-6 | 0.01691 | 0 |
| 1788 | 5.1274E-6 | 5.0403E-6 | 5.0403E-6 | 0.01728 | 0 |
| 1789 | 4.4536E-6 | 4.3937E-6 | 4.3937E-6 | 0.01363 | 0 |

Tab. 6 Von Mises stress error analysis of the MLR method

| NODE ID | CA method | IFU method | Full analysis | Von Mises stress error | |
|---|---|---|---|---|---|
| | | | | CA | IFU |
| 451 | 623.21646 | 771.96421 | 771.96421 | 0.02689 | 0 |
| 452 | 686.68504 | 836.26515 | 836.26515 | 0.02689 | 0 |
| 494 | 1544.71825 | 1588.29902 | 1588.29902 | 0.02744 | 0 |
| 495 | 1541.57486 | 1584.6989 | 1584.6989 | 0.02721 | 0 |
| 817 | 1071.28757 | 1084.77122 | 1084.77122 | 0.01243 | 0 |
| 818 | 1072.06909 | 1081.96295 | 1081.96295 | 0.00914 | 0 |
| 1788 | 674.1351 | 648.47392 | 648.47392 | 0.03957 | 0 |
| 1789 | 630.06725 | 603.10841 | 603.10841 | 0.0447 | 0 |

## 4.3 Bridge

As shown in Fig. 17, a simplified bridge model is considered. The middle of the bridge is subjected to uniformed load $q$ and the dimension of the bridge deck is $L \times D$, where the $L=100mm$, $D=10mm$, the Young's modulus $E=200GPa$, the Poisson's ratio $v=0.3$, and the vertical uniformed load $q=1100mN/mm$. In order to obtain a lightweight design, the shape of the bridge should be modified repeatedly by topology optimization method or others. Assuming the right of Fig. 17 is one of the modified designs with four lighting holes. Then the response of modified structure will be predicted by the MLR method. Considering the symmetry of bridge, a half discrete model has been generated as shown in Fig. 18.

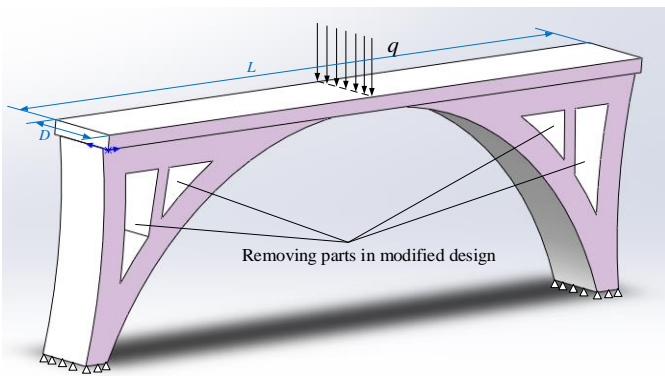

Fig. 17 The modified design of the bridge model

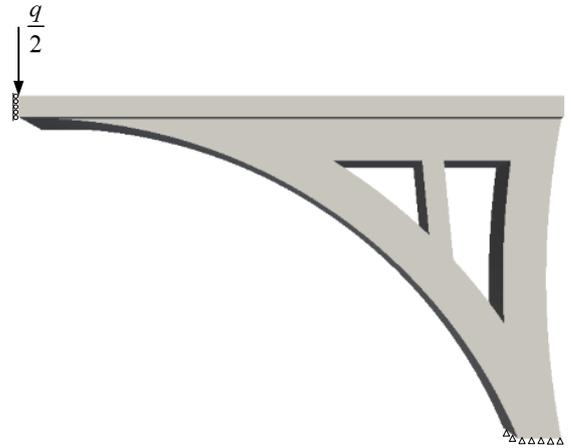

Fig. 18 The half discrete model of the bridge

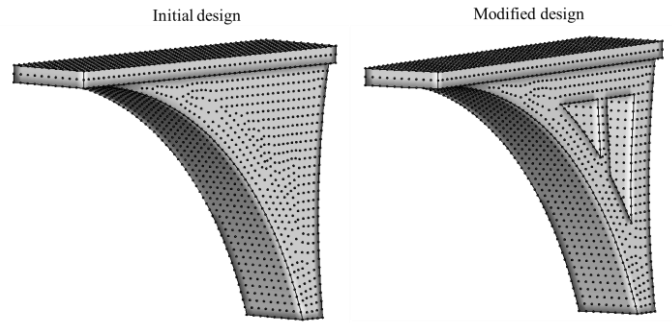

Fig. 19 The initial and the modified nodal distribution of the bridge model

As shown in the left of Fig. 19, 6129 irregularly distributed field nodes are used in the initial structure. To investigate the performance of the MLR, the structure demonstrated in the right of Fig. 19 assumed to be the modified shape. The modified structure is only composed of 5535 nodes, and the percent of reduced DOFs is 9.7% of the original meshless model.

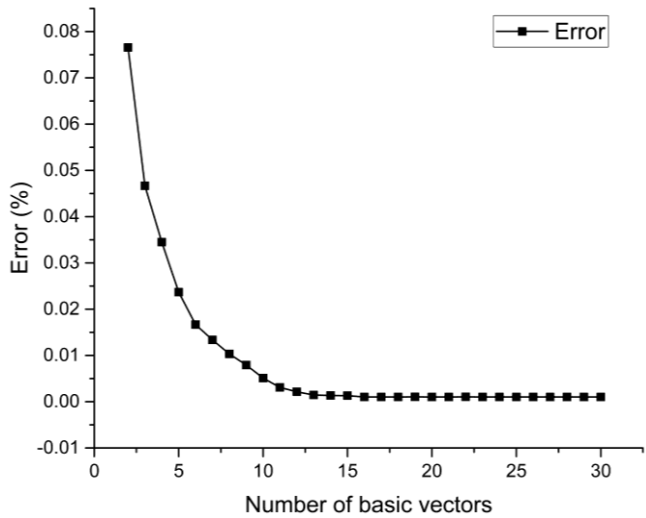

Fig. 20 The error of displacement at different number of basis vectors

As shown in Fig. 20, it can be found that the accuracy of reanalysis solution converges when the number of basis vectors grows up to 15. Sequentially, with more basis vectors, the accuracy of predicted response can't be significantly improved. In this case, 15 is chosen as the number of basic vectors for CA method, and analysis results comparisons between reanalysis and full analysis are illustrated in Fig. 21, Fig. 22, Fig. 23, respectively. Moreover, the displacements of some selected DOFs are listed in Tab. 7. The displacement, strain, stress errors of the CA method are 0.08%, 0.45%, 0.38% respectively while all the errors of the IFU method are 0.

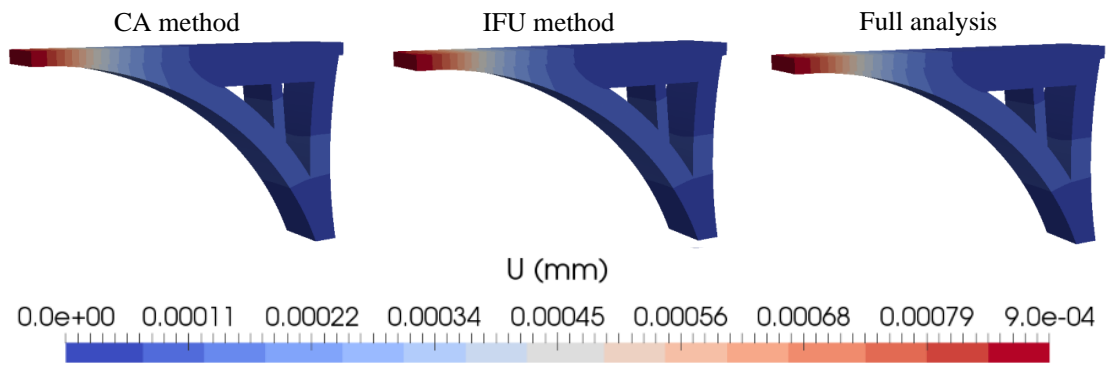

Fig. 21 The displacement result of the MLR and the full analysis methods

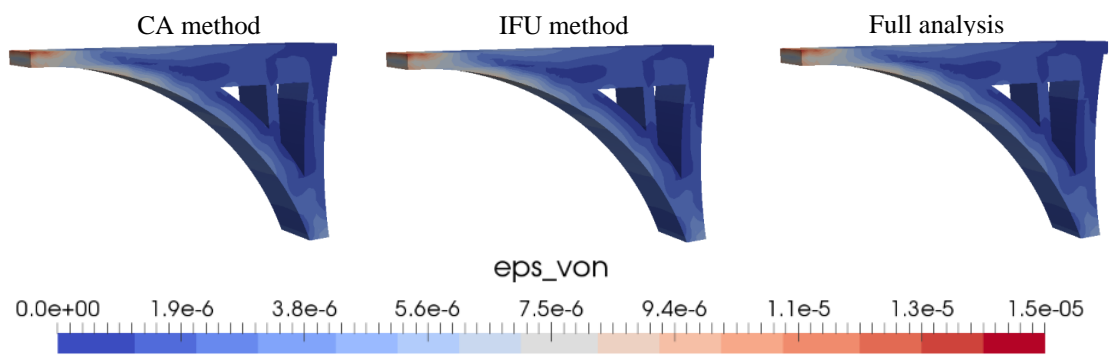

Fig. 22 The von Mises strain result of the MLR and the full analysis methods

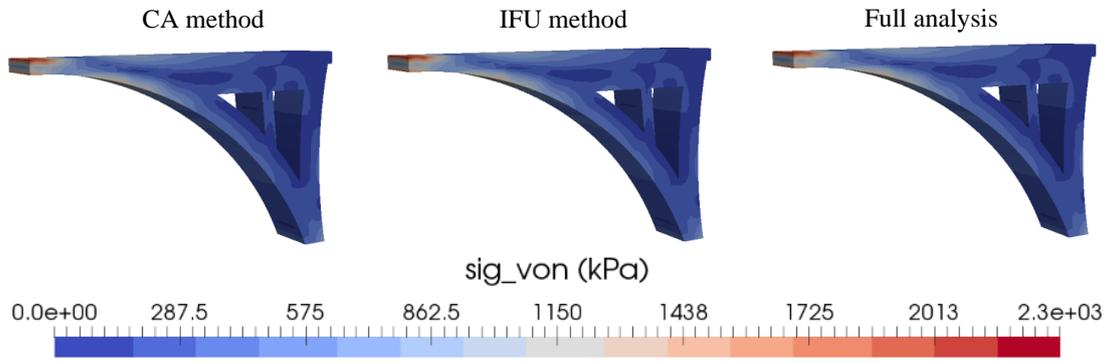

Fig. 23 The von Mises stress result of the MLR and the full analysis methods

Tab. 7 Displacement error analysis of the MLR method

| DOF ID | CA method | IFU method | Full analysis | Displacement error | |
|---|---|---|---|---|---|
| | | | | CA | IFU |
| 131 | -3.1482E-06 | -3.1492E-06 | -3.1492E-06 | 0.000311874 | 1.33E-08 |
| 132 | -0.000797356 | -0.0007974 | -0.0007974 | 5.50875E-05 | 1E-10 |
| 350 | 2.5312E-06 | 2.5417E-06 | 2.5417E-06 | 0.004134 | 9.1E-09 |
| 351 | 1.99659E-05 | 2.01119E-05 | 2.01119E-05 | 0.007259 | 4E-10 |
| 7920 | 5.5871E-06 | 0.000005482 | 0.000005482 | 0.019157159 | 1E-10 |
| 7921 | 6.25185E-05 | 6.24987E-05 | 6.24987E-05 | 0.000316728 | 1E-9 |
| 16066 | 0.000045724 | 4.60102E-05 | 4.60102E-05 | 0.006219645 | 0 |
| 16067 | 4.366E-07 | 4.222E-07 | 4.222E-07 | 0.034057272 | 3.93E-08 |

Tab. 8 Von Mises strain error analysis of the MLR method

| NODE ID | CA method | IFU method | Full analysis | Von Mises strain error | |
|---|---|---|---|---|---|
| | | | | CA | IFU |
| 30 | 1.6865E-06 | 1.6833E-06 | 1.6833E-06 | 0.001915519 | 1E-10 |
| 255 | 6.9695E-06 | 6.9705E-06 | 6.9705E-06 | 0.000149444 | 1E-10 |
| 835 | 2.2692E-06 | 0.00000226 | 0.00000226 | 0.004076097 | 1E-10 |
| 2843 | 2.2043E-06 | 0.000002205 | 0.000002205 | 0.000303044 | 1E-10 |
| 3088 | 2.5427E-06 | 2.5404E-06 | 2.5404E-06 | 0.000929149 | 1E-10 |
| 3124 | 2.2127E-06 | 2.2159E-06 | 2.2159E-06 | 0.001469452 | 1E-10 |
| 4142 | 1.0948E-06 | 1.1048E-06 | 1.1048E-06 | 0.00910551 | 1E-10 |
| 5528 | 1.0132E-06 | 1.0174E-06 | 1.0174E-06 | 0.004159084 | 0 |

Tab. 9 Von Mises stress error analysis of the MLR method

| NODE ID | CA method | IFU method | Full analysis | Von Mises stress error | |
|---|---|---|---|---|---|
| | | | | CA | IFU |
| 30 | 327.4234 | 327.2339 | 327.2339 | 0.000579 | 1E-10 |
| 255 | 808.1318 | 808.2851 | 808.2851 | 0.00019 | 0 |
| 835 | 452.3431 | 450.5761 | 450.5761 | 0.003922 | 1E-10 |
| 2843 | 260.7681 | 260.8149 | 260.8149 | 0.00018 | 1E-10 |
| 3088 | 283.6064 | 283.389 | 283.389 | 0.000767 | 1E-10 |
| 3124 | 226.723 | 227.039 | 227.039 | 0.001392 | 1E-10 |
| 4142 | 167.3127 | 167.6408 | 167.6408 | 0.001957 | 1E-10 |
| 5528 | 116.3698 | 116.5645 | 116.5645 | 0.00167 | 1E-10 |

## 4.4 L-frame

As shown in Fig. 24, a 3D L-frame under uniform load is considered. The top right edge of the L-frame is subjected to uniform load $q$, and the undersurface is fixed. Where the load $q=100mN/mm$, and the Young's modulus $E=200GPa$, the Poisson's ratio $v=0.3$. In order to improve stiffness of the L-frame, a ribbed plate has been added to the right-angle of the L-frame as shown in Fig. 24, and this has been regarded as the modified design.

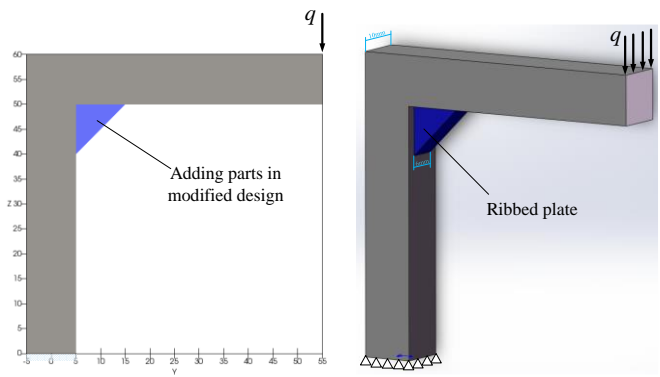

Fig. 24 The modification of the L-frame

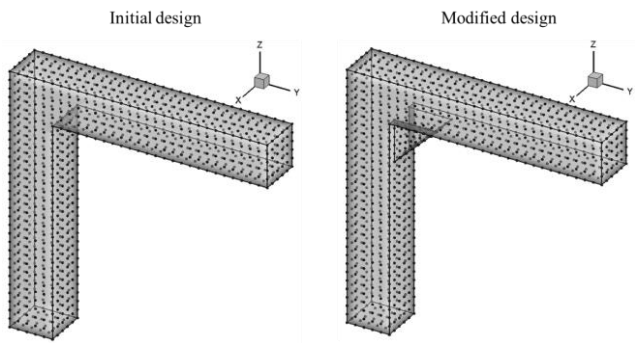

Fig. 25 The initial and the modified nodal distribution of the F-frame

As shown in the left of Fig. 25, 2016 irregularly distributed field nodes are used in the initial structure. To investigate the performance of the MLR, a ribbed plate was added to the L-frame as shown in the right of Fig. 25, and there are 2100 nodes, the percent of adding DOFs is 4.2% of the initial meshless model.

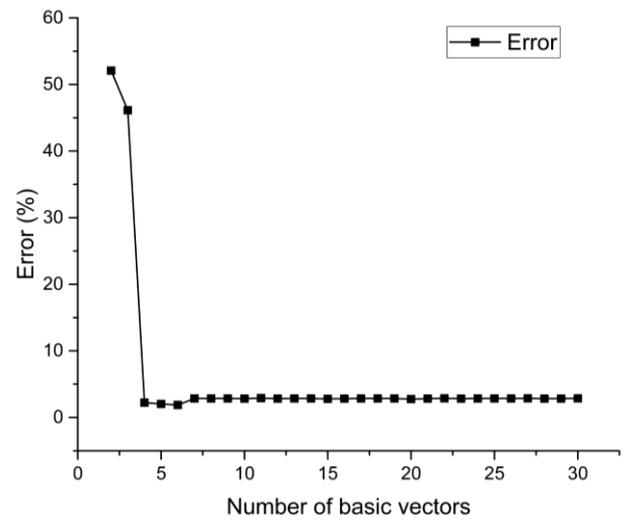

Fig. 26 The error of displacement at different number of basis vectors

As shown in Fig. 26, it can be found that the accuracy of reanalysis solution converges when the number of basis vectors grows up to 7. Sequentially, with more basis vectors, the accuracy of predicted response can't be significantly improved. In this case, 7 is chosen as the number of basis vectors for CA method, and analysis results comparisons between reanalysis and full analysis are illustrated in Fig. 27, Fig. 28, Fig. 29, respectively. Moreover, the displacements of some selected DOFs are listed in Tab.10. The displacement, strain, stress errors of the CA method are 1.87%, 5.6%, 5.2% respectively while all the errors of the IFU method are 0.

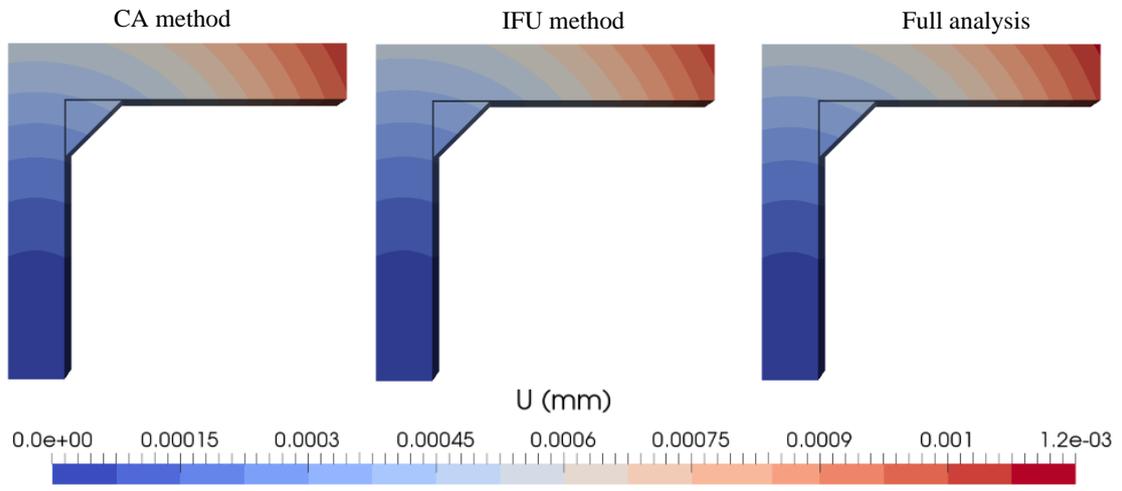

Fig. 27 The displacement result of the MLR and the full analysis methods

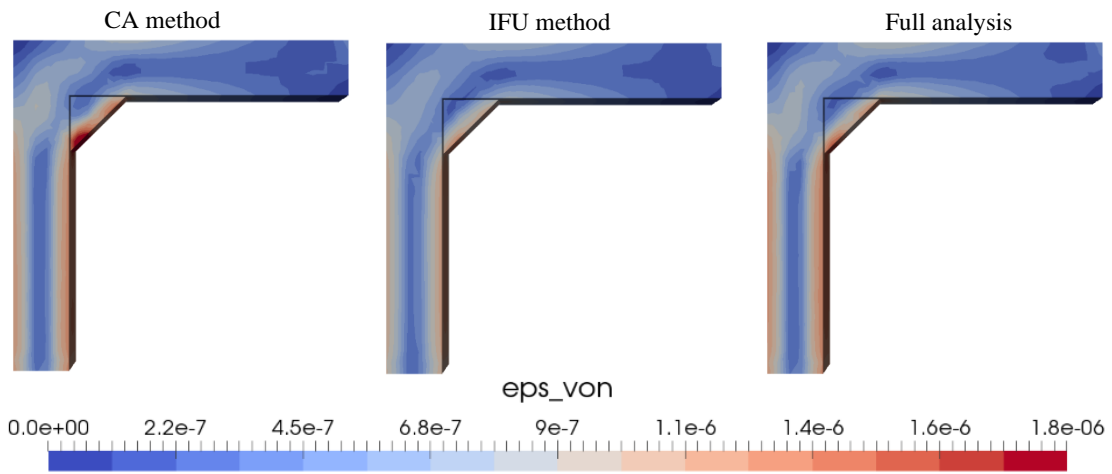

Fig. 28 The von Mises strain result of the MLR and the full analysis methods

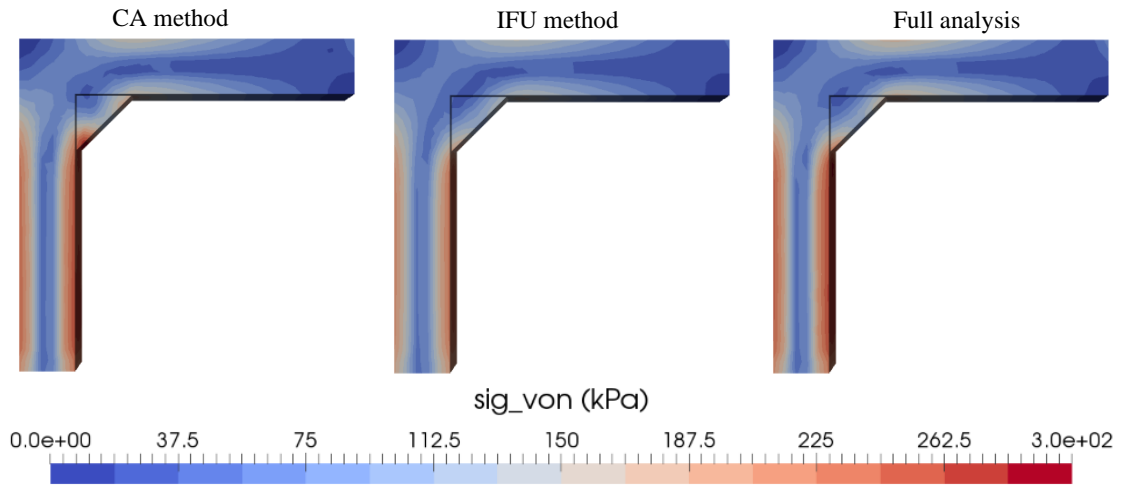

Fig. 29 The von Mises stress result of the MLR and the full analysis methods

Tab. 10 Displacement error analysis of the MLR method

| DOF ID | CA method | IFU method | Full analysis | Displacement error | |
|---|---|---|---|---|---|
| | | | | CA | IFU |
| 29 | 9.2201E-6 | 9.1687E-6 | 9.1687E-6 | 0.00561 | 0 |
| 30 | 1.16594E-5 | 1.17816E-5 | 1.17816E-5 | 0.01037 | 0 |
| 2903 | 1.56476E-5 | 1.56234E-5 | 1.56234E-5 | 0.00155 | 0 |
| 2904 | 1.46668E-5 | 1.48395E-5 | 1.48395E-5 | 0.01164 | 0 |
| 3711 | -5.9957E-6 | -5.9831E-6 | -5.9831E-6 | 0.00211 | 0 |
| 3712 | -9.288E-7 | -9.284E-7 | -9.284E-7 | 3.45795E-4 | 0 |
| 6176 | 3.57369E-4 | 3.64841E-4 | 3.64841E-4 | 0.02048 | 0 |
| 6177 | -9.53544E-5 | -9.57467E-5 | -9.57467E-5 | 0.0041 | 0 |

Tab. 11 Von Mises strain error analysis of the MLR method

| NODE ID | CA method | IFU method | Full analysis | Von Mises strain error | |
|---|---|---|---|---|---|
| | | | | CA | IFU |
| 6 | 1.1546E-6 | 1.1613E-6 | 1.1613E-6 | 0.00579 | 0 |
| 126 | 5.149E-7 | 5.225E-7 | 5.225E-7 | 0.01463 | 0 |
| 418 | 1.3054E-6 | 1.3194E-6 | 1.3194E-6 | 0.01064 | 0 |
| 621 | 3.677E-7 | 3.707E-7 | 3.707E-7 | 0.00822 | 0 |
| 1065 | 3.747E-7 | 3.811E-7 | 3.811E-7 | 0.01666 | 0 |
| 1843 | 1.3226E-6 | 1.3288E-6 | 1.3288E-6 | 0.00464 | 0 |
| 1962 | 8.539E-7 | 8.834E-7 | 8.834E-7 | 0.03341 | 0 |
| 1989 | 1.2031E-6 | 1.2413E-6 | 1.2413E-6 | 0.03071 | 0 |

Tab. 12 Von Mises stress error analysis of the MLR method

| NODE ID | CA method | IFU method | Full analysis | Von Mises stress error | |
|---|---|---|---|---|---|
| | | | | CA | IFU |
| 6 | 204.60153 | 205.98867 | 205.98867 | 0.00673 | 0 |
| 126 | 74.60447 | 77.23895 | 77.23895 | 0.03411 | 0 |
| 418 | 261.05919 | 263.87043 | 263.87043 | 0.01065 | 0 |
| 621 | 73.44396 | 74.12374 | 74.12374 | 0.00917 | 0 |
| 1065 | 67.5001 | 67.48854 | 67.48854 | 1.71261E-4 | 0 |
| 1843 | 264.46262 | 265.71453 | 265.71453 | 0.00471 | 0 |
| 1962 | 170.68602 | 176.60388 | 176.60388 | 0.03351 | 0 |
| 1989 | 240.61834 | 248.24881 | 248.24881 | 0.03074 | 0 |

### 4.5 Accuracy and efficiency comparison

Four numerical examples have been tested in this section, these cases include large and small modification, reduce and increase nodes, 2D and 3D problems. It can be found that the accuracy of CA method is high for both 2D and 3D problems, even for large modification. Meanwhile, the IFU method can obtain the exact response of the modified structure, there are almost no errors for both four cases. Moreover, the accuracy of reducing nodes modifications is much higher than adding nodes modifications by using CA method and the accuracy of stress-strain results is lower than displacement results.

Tab. 13, the computational time of stiffness calculation and solution by different reanalysis methods under different stiffness matrix updating strategies are listed. It can be found that the local stiffness matrix updating strategy saves much stiffness calculation cost than traditional global

To investigate the performance of the MLR method, the CPU running time which cost by the full analysis and MLR method has been recorded and all the simulations were performed on an Intel(R) Core(TM) i7-5820K 3.30GHz CPU with 32GB of memory within MATLAB R2016b in x64 Windows 7. As shown in stiffness matrix updating strategy, especially for local modification problems. It also can be found that the efficiency of reanalysis methods is much higher than the full analysis, more detail is described in Fig. 30 and Fig. 31.

Tab. 13 The computational time of the relative components

| Numerical examples | CPU time (*s*) | | | | |
|---|---|---|---|---|---|
| | Local updating strategy | Global updating strategy | CA reanalysis | IFU reanalysis | Full analysis |
| **Rectangular plate** | 61.14 | 214.53 | 0.14 | 1.34 | 1.62 |
| **Support bracket** | 6.12 | 60.93 | 0.07 | 0.04 | 0.83 |
| **Bridge** | 1640.81 | 2617.7 | 2.86 | 3.29 | 23.65 |
| **L-frame** | 292.71 | 426.64 | 0.26 | 0.32 | 2.37 |

Moreover, in order to fully investigate the efficiency of the MLR method, two representative cases which include large modification (the change ratio is about 34%) and small modification (the change ratio is about 1.5%) have been calculated by MLR method under different computational scales. The log-log plots of comparison results are shown in Fig. 30 and Fig. 31 and the error analysis was also shown.

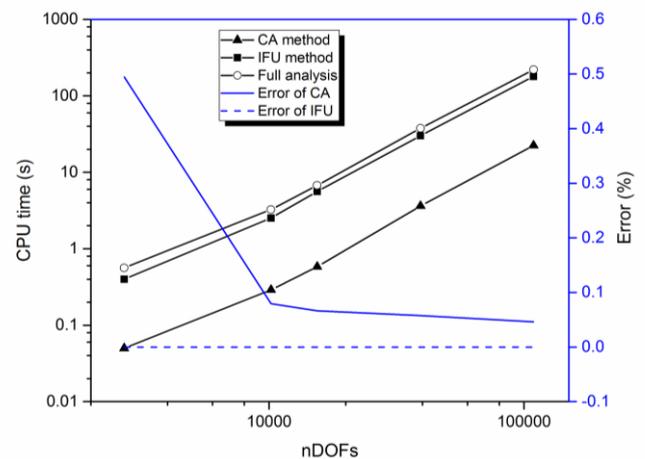

Fig. 30 The comparison of computational cost (large modification)

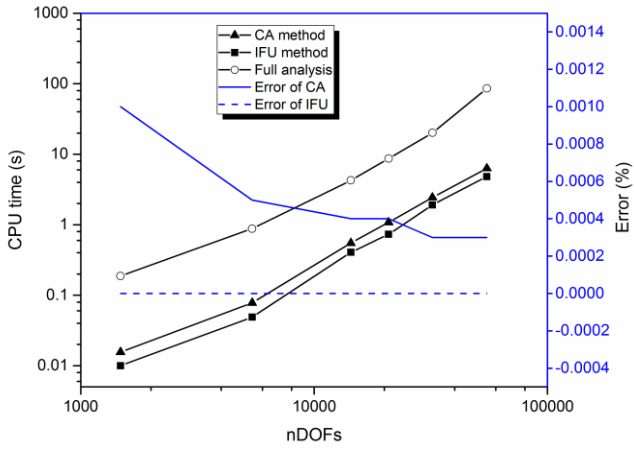

Fig. 31 The comparison of computational cost (small modification)

It is obvious that the efficiency of the IFU method is much higher than the full analysis for the small modification, but for the large modification, the efficiency of IFU method should be largely reduced while the CA method behaves much better. In addition, the accuracy of CA method should be improved as the increase of number of DOFs while the IFU method is an exact reanalysis method. Although the accuracy of CA method is lower than IFU method, but the CA method is much more efficient than IFU method when meeting large modification problems. In a word, the CA is a universal method for both large and small modifications while the IFU method is much more suitable for small modifications.

## 5 Conclusions

Compared with other FE-based reanalysis methods, the meshless-based reanalysis method is easier to be implemented because only points should be added or removed while a structure is modified without considering the connection of nodes. By using the MK method to construct stiffness matrix, the specified essential boundary conditions can be easily implemented due to the property of Kronecker's delta function that Kriging interpolation procedure possesses. However, because more nodes in the influence domain are involved in constructing shape function, more relative nodes should be considered in building the modified stiffness matrix. Therefore, a local strategy is suggested. By this strategy, only the nodes inside the influence domain are used to construct local stiffness matrix rather than all nodes. Furthermore, considering the expensive computational cost of meshless methods, the advantage in term of efficiency of the meshless-based reanalysis is more obvious.

The effect of the number of basis vectors on analysis results has been discussed in this study, and the strain and stress formulations based on MLR are also given to make a comparison of accuracy between reanalysis and full analysis. Four numerical examples have shown that the accuracy of the MLR method is available even for large modification problems and this method can save much computational cost. Moreover, this study not only made comparisons of displacement, but also made comparisons of strain and stress, and the result shows that the accuracy of stress-strain results is available.

In summary, the MLR method is a high-efficiency method with nice accuracy both in displacement, strain and stress.

However, further research is still needed to improve the accuracy of the meshless-based reanalysis method for the large deformation problems. It should also be extended to dynamic problems to fully show the advantages, such as crack propagation.

**Acknowledgements** This work has been supported by Project of the Program of National Natural Science Foundation of China under the Grant Numbers 11572120 and 61232014.